\documentstyle[aps,epsf,eqsecnum]{revtex}

\begin{document}
%\twocolumn[\hsize\textwidth\columnwidth\hsize\csname @twocolumnfalse\endcsname

\draft

\title{
Role of Conduction Electrons in the ortho-KC$_{60}$ Polymer
}
\author{T. Ogitsu, K. Kusakabe, S. Tsuneyuki and Y. Tateyama}
\address{
Institute for Solid State Physics, University of Tokyo, Roppongi,
Minato-ku, Tokyo 106, Japan\\
}
\date{\today}
\maketitle

\begin{abstract}
We present first-principles band calculations as well as structural
optimization of the orthorhombic K$_1$C$_{60}$ polymerized phase. We
found three-dimensional inter-fullerene bonding/anti-bonding
characters consisting of $t_{1u}$ molecular orbitals with $p$-like
symmetry in the conduction bands. The formation of the four-membered
ring connecting the fullerenes lifts the $p_x$ band from the
continuous $p_y$-$p_z$ band, where the z-axis is parallel to the
chain. The asymmetry between $p_x$ and $p_y$ may play an important
role in
binding the chains with the proposed rotational configuration\cite{Stephens}.
%
%The inter-fullerene-bonding character in the lowest
%conduction band shows that the doped electrons not only help to form
%the one dimensional chain itself, but condense the chains in the
%body-centered-orthorhombic crystal.
\end{abstract}
%]
%
\narrowtext The recent discoveries of new alkali-doped fullerene
polymers\cite{Stephens}\cite{Chauvet}\cite{Prassides}
\cite{Bendele}\cite{Oszlanyi}
show diversity in the geometries of the interconnections of fullerenes.
%The two dimensional\cite{Oszlanyi} as well as one dimensional
%polymers\cite{Stephens}\cite{Chauvet}\cite{Prassides}\cite{Bendele} have
%been reported.
An impressive property of this polymer family is that the charge state of
the fullerene seems to govern the polymer geometry. The A$_1$C$_{60}$
(A=K, Rb)\cite{Stephens}\cite{Chauvet} and the Na$_2$RbC$_{60}$
polymers\cite{Prassides}\cite{Bendele} have one-dimensional
interconnections, while the Na$_4$C$_{60}$ polymer has two-dimensional 
interconnections. 
Even in the case of the one-dimensional polymers, we can
find a remarkable difference between the A$_1$C$_{60}$ (A=K, Rb) and the
Na$_2$Rb'C$_{60}$ polymers; the former is known to have two
covalent bonds between the fullerenes\cite{Stephens}, 
while the latter is
%deduced to have one covalent bond from its larger
%inter-molecular separation compared with the former
%one\cite{Prassides}\cite{Bendele}.  
deduced
from its larger
inter-molecular separation\cite{Prassides}\cite{Bendele}  
to have one covalent bond. 
These dramatic differences in bonding which are controlled
by the charge states give rise to the question as how the excess
electrons on the fullerenes affect the polymer geometry.

In this paper, we present  a first-principles electronic structure
study of the ortho-KC$_{60}$ polymer to elucidate how the electronic
states of the conduction bands are related to its one-dimensional polymer
geometry. Although electronic band structure calculations of A$_1$C$_{60}$
have been reported by two other groups\cite{Tanaka1}\cite{Erwin}, to our
knowledge, this is the first quantitatively reliable calculation on the
system with a realistic atomic configuration. The calculation method is based
on the density functional
theory within the local density approximation
(DFT-LDA)\cite{method}. A plane-wave basis expansion and optimized
pseudopotentials in separable form were adopted in this simulation.

%\vspace{\baselineskip}

Structural optimization is performed on the KC$_{60}$ polymer, which
has an orthorhombic unit cell containing two C$_{60}$ molecules and
two K atoms. The cell parameters are fixed at the experimental
values\cite{Stephens}. 
%As for the initial geometry, 
Initially,
the C$_{60}$
molecules are placed at the body-centered positions, in
which the atomic configuration of an isolated $I_h$ C$_{60}$ 
molecule is used.
We choose the initial orientational configuration of the fullerenes as the one
proposed by Stephens {\em et al.}\cite{Stephens}, {\em i.e.}, 
the unit cell is
taken to be simple orthorhombic. The structural
optimization is then performed without any symmetry constraint. To obtain
as precise a geometry as possible, we have checked the convergency of the
geometry as a function of the cutoff energy for 
plane waves ranging from 30Ry to 80Ry with uniform sampling of 
either one ($\Gamma$) or four
$k$-points. The geometry is relaxed until all the
forces acting on the atoms become less than $5\times 10^{-4}$
[Hartree/au]. 
%Consequently, the 
Covalent bonds spontaneously appeared  %in
between the
C$_{60}$ molecules during the optimization process, 
resulting in a  one-dimensional
chain structure.

The calculated bond lengths are displayed in TABLE~1 along with the
experimental values.
The calculated values are consistent with the experimental values within 
the limit of error, 
with the largest discrepancy occurring in the C1-C1
intra-fullerene distance. Refinement of the experimental value (1.74\AA)
is desired, since it seems 
%too %too is quite strong here
quite
large compared with the typical C-C bond lengths known thus far.
%The consistency between the calculated values and
%the experimental ones can be seen except for the C1-C1 intra-fullerene
%distance. The error bars of the experimental values are so large
%that refinement is desired.

%%%%%%%%%%

Let us discuss the geometrical character of the KC$_{60}$ polymer from
the viewpoint of the coupling between the conduction electrons and the
deformation of the C$_{60}$ molecule. It would be natural to speculate
that the character of the LUMO state of isolated C$_{60}$ would be
reflected in the conduction band. Therefore, we shall note the
symmetry of the LUMO state of isolated $I_h$ C$_{60}$ before
explaining the resultant structure.

The LUMO of the $I_h$ C$_{60}$ molecule are threefold-degenerate $t_{1u}$
states with $p$-like symmetry. The coordinate axes for these
$p$-like ($t_{1u}$) states can be mapped on the three orthogonal $C_2$
rotational pivots on the $I_h$ C$_{60}$ molecule.

If we now go back to the calculated polymer structure, we note that
the chain direction is parallel to one of the $C_2$ rotational pivots. 
Therefore, if we take the chain direction as the z-axis of the
$t_{1u}$ states, the other two coordinate axes are automatically
determined: one (x-axis) is on the plane which includes the
four-membered ring of the inter-molecular connection and the other
(y-axis) is normal to the above two. We then measure the diameters of
the polymerized fullerene along these coordinate axes. Interestingly,
even the two diameters normal to the chain are significantly different
from each other: they are 6.98\AA \ and 6.83\AA \ along the x- and
y-axes, respectively. As is expected, the diameter along the z-axis is
elongated to 7.48\AA.

We emphasize here that this ellipsoidal distortion may be plausibly
associated with the Jahn-Teller effect; however, as will be clarified
later in this paper, the relative positions of the $t_{1u}$ levels in
the polymer are instead determined by the crucial change in the local
electronic states around the four-membered ring connecting the
fullerenes as well as the strength of the inter-molecular interaction.

%\vspace{\baselineskip}

The electronic states of this orthorhombic polymer crystal have been
investigated using the first-principles band calculation method. The
householder diagonalization method with a plane-wave expansion is used
to maintain the symmetry of the eigenstates as precisely as
possible. The cutoff energy for the plane waves is reduced to 25Ry to
save on the computational cost. The number of basis functions is then
reduced to 25000. We note here that the convergence of the band
structure has been confirmed for both diamond and potassium crystals
with this cutoff energy, though it is inadequate for the structural
optimization. Because the diagonalization of a matrix with a dimension
of 25000 requires a huge amount of CPU time even for the newer
parallel supercomputers, we did not perform fully a self-consistent
diagonalization calculation but instead obtained the eigenvalues and
the eigenstates through the following two steps. First, the effective
potential for the Kohn-Sham equation was calculated self-consistently
by means of conjugate gradient minimization of the total energy, in
which 75 uniform $k$-points in the first Brillouin zone were
sampled. We then diagonalized the matrix calculated from the given
effective potential.

The calculated band structure is shown in FIG. 2. Although the band
dispersion along the $\Gamma$-H line (parallel to the chain) is slightly
larger than that along the $\Gamma$-M line (normal to the
chain), the band structure indicates essentially three-dimensional character.
This is qualitatively consistent with previous
calculations\cite{Tanaka1}\cite{Erwin}, although the band profile is
considerably different. % from theirs.  from those?
Empirical tight-binding calculations
by Tanaka {\em et al.}\cite{Tanaka1} gave a similar band structure along the
$\Gamma$-M line, while the $\Gamma$-H dispersion was not fully represented
by the simplified model Hamiltonian. 
In a different paper, Tanaka {\em et al.} also pointed out
that the dispersion of the conduction band is very sensitive to the
orientational configuration of the chains\cite{Tanaka2}. This is probably
why the discrepancy with the DFT-LDA calculation for RbC$_{60}$ by Erwin 
{\em et
al.}\cite{Erwin} is much worse: they used a model structure in which all the
chains have the same orientation so that the unit cell includes only one
A$_1$C$_{60}$ unit, while in the Reitveld model\cite{Stephens} and therefore 
in
the resultant structure of the present calculation, the adjacent chains are
rotated by 90$^\circ$ from each other.

%Tanaka et al. have performed an empirical tight-binding
%calculation\cite{Tanaka1}. Their band dispersion along the $\Gamma$-M
%line is similar to the present result, while the dispersion along
%the $\Gamma$-H line is remarkably different; even
%the lowest and the fourth lowest bands at the $\Gamma$-point cross
%each other in our band structure, while only the second and the third
%lowest bands intersect each other in their band structure.

%Erwin et al. have calculated the band structure of the RbC$_{60}$
%polymer using the first-principles DFT-LDA method\cite{Erwin}. The
%discrepancy from ours will come from their choice on the orientational
%configuration of the chains; they used a body-centered orthorhombic
%cell in which all the chains have been rotated by 45 degree from the
%Reitveld model\cite{Stephens}. As it has been shown by Tanaka et al.,
%the dispersion of the conduction band is very sensitive to the
%orientational configuration of the chains\cite{Tanaka2}. Therefore,
%the electronic structure analysis using the Reitveld model is needed.

%\vspace{\baselineskip}

Next, we would argue the character of the conduction band. The lower
six conduction bands seem to be categorized 
from their profiles along $\Gamma$-M line
as three sets of two pairs
as follows (FIG. 1.):  
looking at the energy levels at
the $\Gamma$-point, the lowest and the third lowest, the second and
the fourth lowest, the fifth and the sixth lowest, can be recognized
as pairs.

Direct inspection of these wavefunction ensures the above
categorization. We have examined the iso-surfaces of these six
wavefunctions at the $\Gamma$-point. Those belonging to the 
lowest and  the third lowest
are displayed in FIG. 2 (a) and (b), respectively.

We can clearly see the bonding and the anti-bonding characters of the
wavefunctions between the polymer chains in FIG. 2. (a) and (b),
respectively. Furthermore, we can see the $t_{1u}$ character surviving
in these states, {\em i.e.} the $\pi^*$ character of the C=C double
bonds and the nodal plane which includes the four-membered ring
connecting the fullerenes. There is almost no evidence of
hybridization with the other states.  The $t_{1u}$ symmetry and the
bonding/anti-bonding characters as described above are also seen in
the other four eigenstates, though the $t_{1u}$ mark in the sixth
eigenstate is weakened; the second (see FIG. 3.) and the fourth lowest
eigenstates at the $\Gamma$-point have a nodal plane normal to the
chain, while the fifth (see FIG. 4.)  and the sixth lowest eigenstates
have a nodal plane normal to the above two nodal planes.

We can now understand that the wavefunctions of the conduction
bands of the ortho-KC$_{60}$ polymer are, at least at the $\Gamma$-point,
linear combinations of the $t_{1u}$ molecular orbitals (MOs) as follows,
\begin{equation}
\Psi_{p_i,\Gamma} = \frac{1}{\sqrt{2}}(\phi_{p_i} \pm
\phi'_{p_i}),\mbox{\hspace{1em}}i=x,y,z.
\end{equation}
Here, $\phi_{p_i}$ and $\phi'_{p_i}$ denote the MOs belonging to 
the first and
the second fullerenes in the unit cell, respectively. Note that the x-axis
for the first fullerene is taken to be parallel to the y-axis for the second
fullerene, since the coordinate axes of these $p$-like states for each
fullerene have been taken independently so that the x-axis includes the
four-membered ring connecting the fullerenes.

Due to the polymerization, the threefold degenerate 
$t_{1u}$ states split into three different
levels, $E_{p_x}$, $E_{p_y}$, and $E_{p_z}$. 
These levels then constitute the inter-chain
bonding/anti-bonding states, and the lowest state which has the
inter-chain bonding character will be occupied. 
%As it can be seen as
%the clear inter-chain bonding character in FIG. 2. (a), the inter-chain
%interaction is large enough to form the conduction band with the three
%dimensional dispersion.
Clear inter-chain bonding character can be seen in FIG. 2. (a), 
where the inter-chain
interaction is large enough to form a conduction band with 
three-dimensional dispersion.

%\vspace{\baselineskip}

We note here that the occupied levels associated with the
four-membered ring connecting the fullerenes sink below the
energy gap and so have not appeared in the above discussions. This
may be confounding because alkali doping is known to be crucial
to the synthesis of the ortho-AC$_{60}$. 
Below, we argue as to how the formation
of the covalent connections couples with the profile of the
conduction band.

We now focus on the change in the local electronic states around the
carbon atoms forming the four-membered ring. Due to the polymerization, the
local electronic states are converted from $sp^2$ to $sp^3$. This means that
the
$\sigma^*$ and $\pi^*$ states of the $sp^2$ re-hybridize into
the $\sigma^*$ states of the $sp^3$. We emphasize here that 
$\phi_{p_x}$ has an amplitude around the ring (see FIG. 4), while 
$\phi_{p_y}$ and $\phi_{p_z}$ do not. This means that only the
levels which are functions of $\phi_{p_x}$ will be affected by the
re-hybridization. As can be seen from FIG. 1., the $p_x$
band is lifted about 0.7eV from the $p_y$--$p_z$ band, and forms a
continuous band with the higher energy states.

%%%%%%%%%%%%%%%%%%%%%%%%%%%%%

To confirm above the scenario in which the re-hybridization lifts only
the $E_{p_x}$ level, we have performed first-principles MO
calculations on two different C$_{60}$ molecules having geometry
equivalent to the polymer using Gaussian~94\cite{Gaussian}. One is the
bare C$_{60}$ molecule and the other is the C$_{60}$ molecule with the
two carbon atoms which belong to the four-membered ring terminated
with hydrogen bonds.  In these calculations, Becke's 3 parameter
hybrid method\cite{Becke} with Perdew's gradient-corrected correlation
functional\cite{Perdew} (B3P86) and a 6-31G basis set were used.
The bare deformed fullerene shows a too-low $E_{p_x}$ level, with
$E_{p_x}= -4.6eV < E_{p_z}= -4.2eV < E_{p_y}= -4.1eV$, while hydrogen
termination serves to lift the $E_{p_x}$ level, giving levels
comparable to the band calculation, with $E_{p_z}= -4.0eV < E_{p_y}= -
3.9eV < E_{p_x}=-2.9eV$. Thus, we conclude that the formation of the
four-membered ring connecting the fullerenes raises the $p_x$ band
from the continuous $p_y$--$p_z$ band.
In the latter MO calculation, the sequence, $E_{p_z} < E_{p_y}$, is
opposite from the $\Gamma$-point levels of our band calculation,
but the difference is small in both cases. This is due to the
crystal field effect, which is obserbed as the difference in the
strength of the inter-fullerene interaction; in  
$\Psi_{p_y,\Gamma}$, intra-chain  as well as inter-chain bonding
character can be seen between the fullerenes, while the no
intra-chain bonding character can be seen in $\Psi_{p_z,\Gamma}$
(see FIG. 3).

%In summary, from the MO calculations, we have found that $E_{p_x}$ is
%lifted considerably by the formation of the four-membered ring
%connecting the fullerenes, while from the band calculations, we have
%found that $E_{p_y,\Gamma}$ becomes lower than $E_{p_z,\Gamma}$ due to
%the intra-chain interaction seen in $\Psi_{p_y,\Gamma}$\cite{Ogitsu}.

%\vspace{\baselineskip}

Finally, we argue as to how the orientational configuration of
the chains could be restricted as was reported 
experimentally\cite{Stephens}.
There may be two key factors, one of which is
the symmetry of the $t_{1u}$ molecular orbitals, the other being 
the rise of
the $E_{p_x}$ level above $E_{p_y}$ and $E_{p_z}$  as seen in our results.

Nikolaev {\em et al.} calculated the $t_{1u}$ level splitting in the
FCC A$_1$C$_{60}$ crystal\cite{Nikolaev} as the function of the
orientational configuration of the fullerenes by model
calculations in which the symmetry of the $t_{1u}$ state as well as
the charge
state of the fullerene are taken into account. They found  that the two
configurations which correspond to the local maxima in the energy
splittings can be extended to 
represent the ortho-A$_1$C$_{60}$ (A=,K,Rb)
and monoclinic ${\rm A_2A^{\prime}C_{60}}$
(A=Na,A$^\prime$=K,Rb) polymers. %, respectively.
%Even in the orthorhombic A$_1$C$_{60}$ polymer, it was found that the
%two orientational configurations of the chains give the local minima
%in the band width by Tanaka {\em et al.} using the empirical tight binding
%method\cite{Tanaka2}.
Tanaka {\em et al.,} using the empirical tight binding
method\cite{Tanaka2}, found that even in the orthorhombic A$_1$C$_{60}$
polymer, two orientational configurations of the chains give local
maxima in the band width.

Although the threefold degenerate $t_{1u}$ eigenfunctions have
$p$-like symmetries, they cannot be represented by spherical
harmonics with $\ell=1$. Owing to the icosahedral symmetry of the
fullerene, they must be represented as the linear combinations of
$Y_{\ell m}$ with $\ell=5$\cite{Nikolaev}. The above studies 
suggest that the symmetry of the $t_{1u}$ state is the major factor in
determining the geometry of the ortho-A$_1$C$_{60}$ polymer. We then
have to question as to why, if the model calculations show comparable
energies for both local minima, A$_1$C$_{60}$ chooses an
angle of 45 degrees.

%We now speculate that 
We can speculate that
the asymmetry between the x-direction and the
y-direction around the chain causes the energy difference between the
orientational configurations. The considerable rise of the 
%$E_{p_x}$
${p_x}$
band from the continuous $p_y$--$p_z$ band, or the difference in the
diameters along the x-axis and along the y-axis may be the reason for
the preference of the observed orientational
configuration\cite{Stephens}.

Further first-principles calculations of other
orientational configurations, or of polymers with
different charge states, are required to clarify the role of the conduction
electrons in determining the polymer geometry.

%\vspace{\baselineskip}

In conclusion, we have performed semi-SCF DFT-LDA band calculations
as well as first-principles geometry optimization of the
ortho-KC$_{60}$ polymer. The simulations were performed using 
cell dimensions
equivalent with experiment. The resulting band has a three-dimensional
dispersion which has a considerably different profile from those
previously calculated\cite{Tanaka1}\cite{Erwin}.
We found that the lower six conduction eigenstates 
at the $\Gamma$-point are represented as the linear combination
of the $t_{1u}$ MOs. The rise of the $p_x$ band is found to be due to
the local conversion of the electronic state, from $sp^2$ to $sp^3$,
by the formation of the four-membered ring connecting the
fullerenes. The conduction electrons will assist to form the one-dimensional 
chain as well as to bind the chains in the
orthorhombic cell with this peculiar orientational configuration.

The authors thank Professor Iwasa for stimulating
discussions. T.O. thanks Dr. Suzuki for helpful suggestions. T.O. also
thanks Dr. T. M. Briere for critical readings of this manuscript. This
work was performed on Fujitsu VPP500 supercomputers at the super
computer center, Institute for Solid State Physics, Univ. of Tokyo and
at the computer center, Nagoya University, the VPP700 at RIKEN and at
the computer center, Kyushu University, and on the Hitach SR2201 at
the computer centre, University of Tokyo. This work was conducted as
part of the JSPS Research for the Future Program in the Area of
Atomic-Scale Surface and Interface Dynamics.

%\newpage
\begin{table}[h]%1
  \begin{tabular}[h]{lll}% \hline
     & Experiment & This work \\ \hline
    Lattice parameter [\AA] &  9.109(5) & ------- \\
     &  9.953(5) & ------- \\
     &  14.321(5) & ------- \\
    C1-C1 inter-fullerene & 1.65$\pm$ 0.20 & 1.62 \\
    C1-C1 intra-fullerene & 1.74$\pm$ 0.20 & 1.59 \\
    C1-C2, C1-C3 & 1.50$\pm$ 0.20 & 1.51 \\
%    \hline
  \end{tabular}
\caption{The calculated structural parameters and the experimental
  values. The notation is the same as Stephens 
  {\em et al.}\protect\cite{Stephens}}
\end{table}
%\newpage

\begin{figure}[htbp]
%  \begin{center}
%    \leavevmode
%    \epsfile{width=8cm,file=KC60band.epsi}
%  \end{center}
\caption{Band structure of the ortho-KC$_{60}$ polymer. 
%The lower six conduction bands are drawn above the energy gap. 
The $\Gamma$-H
direction is parallel to the chain and the $\Gamma$-M direction is
normal to the chain. The vertical axis denotes the energy in eV. The origin
placed at the fermi level. Note: the fermi level is not determined
self-consistently; it is evaluated as the energy where the filling
becomes  half of the full occupation number during preparation of the
effective potential by the CG method.}
\end{figure}
%\newpage

\begin{figure}[htbp]
%  \begin{center}
%    \leavevmode
%    \epsfile{width=8cm,file=bondc.ps}\\
%    (a) \\
%    \vspace{0.5cm}
%    \epsfile{width=8cm,file=anti-bondc.ps}\\
%    (b)
%  \end{center}
\caption{The iso-surfaces of the lowest, (a), and the third lowest,
  (b), eigenstate wavefunctions at the $\Gamma$-point are drawn
  with the atoms represented by a ball and stick model in the
  orthorhombic unit cell. The yellow balls denote the carbon atoms, and
  the white balls denote the potassium atoms. The viewpoint is parallel
  to the chain direction.  The iso-level is set to 1/5 of the
  maximum absolute amplitude. The yellow and the blue curved surfaces
  denote the iso-surfaces of the positive and negative values
  of the wavefunction, respectively. }
\end{figure}

%\newpage

\begin{figure}[htbp]
%  \begin{center}
%    \leavevmode
%    \epsfile{width=8cm,file=second_side.eps}\\
%  \end{center}
\caption{The iso-surfaces of the second lowest eigenstate
  wavefunction at the $\Gamma$-point is drawn
  with the atoms represented by a ball and stick model in the
  orthorhombic unit cell. The viewpoint is normal to the chain direction.}
\end{figure}

\begin{figure}[htbp]
%  \begin{center}
%    \leavevmode
%    \epsfile{width=8cm,file=fifth_cont.eps}\\
%  \end{center}
\caption{The contour plot of the fifth lowest eigenstate
  wavefunction at the $\Gamma$-point is drawn with the atoms 
  represented by a
  ball and stick model in the orthorhombic unit cell. The blue and 
  red areas correspond to the negative and positive values of the
  wavefunction, respectively. The viewpoint is normal to the chain
  direction. The contour plane includes the four-membered ring
  connecting the fullerenes.}
\end{figure}

\end{document}